\documentclass[a4paper,11pt,msmath,amssymb,superscriptaddress,floatfix,nofootinbib]{article}
\usepackage{jheppub} 
\usepackage{lineno}
\usepackage{exscale}
\usepackage{relsize}
\usepackage{graphicx}
\usepackage{dcolumn}
\usepackage{bm}
\usepackage{multirow}
\usepackage{CJK}
\usepackage{diagbox}
\usepackage{subfigure}
\usepackage{footmisc}
\usepackage{footnote}
\usepackage{xcolor}
\usepackage{amsmath}
\usepackage{array}
\usepackage{textcomp}
\usepackage{wasysym}
\usepackage{float}
\usepackage{subcaption}
\usepackage{caption}
\usepackage{amsfonts}
\usepackage{amssymb}
\usepackage{bbold}
\usepackage{epstopdf}

\usepackage{hyperref} 
\usepackage{url}      

\usepackage{natbib}

\hypersetup{colorlinks=true,citecolor=blue,anchorcolor=red,menucolor=red,linkcolor=red,filecolor=true,
runcolor=red,urlcolor=blue,frenchlinks=true}

\arxivnumber{} 

\title{Normalization of partial wave CP asymmetries in three-body decays of heavy hadrons}

\author[a,c]{Jing-Juan Qi,}

\author[c]{Zhen-Yang Wang,}

\author[b]{Zhen-Hua Zhang,\footnote{Corresponding author}} 

\author[d]{and Xin-Heng Guo\footnote{Corresponding author}}

\affiliation[a]{College of Information and Intelligence Engineering, Zhejiang Wanli University, Zhejiang 315101, China,}
\affiliation[b]{School of Nuclear Science and Technology, University of South China, Hengyang, 421001, Hunan, China}
\affiliation[c]{Physics Department, Ningbo University, Zhejiang 315211, China}
\affiliation[d]{School of Physics Science and Technology, Kunming University, Kunming 650214, China}

\emailAdd{jjqi@mail.bnu.edu.cn}
\emailAdd{wangzhenyang@nbu.edu.cn}
\emailAdd{zhangzh@usc.edu.cn}
\emailAdd{xhguo@bnu.edu.cn}

\abstract{
CP violation in hadronic multi-body decays has been extensively studied, and the experimental breakthrough in the heayy baryon sector was made recently. 
Partial-Wave CP Asymmetries (PWCPAs) in multi-body decays of heavy hadrons, which although provide us with more interference information, suffer from the
normalization problem, as is pointed out in this paper.
We propose a novel solution to this problem.
We introduce a set of extra factors to rescale the PWCPAs to the proper sizes.
Instead of determining the set of factors according to the normalization requirement, we demand that all the PWCPAs have the same statistical errors.
In this way, we obtain a set of quasi-normalized PWCPAs, in the sense that they are close to the ideal normalized ones.
As an application, we perform an analysis of PWCPAs in the decay channel $B^\pm\to\pi^+\pi^-\pi^\pm$. We focus on the phase space region where the invariant mass of the $\pi^+\pi^-$ pair varies around the vector resonance $\rho^0(1450)$. Based on the data of the LHCb collaboration, the interference patterns among the resonances $\rho^0(1450)$, $f_2(1270)$, and $f_0(1500)$ and their contributions to  the quasi-normalized PWCPAs are analyzed.
The analysis indicates that the quasi-normalized PWCPAs can avoid potential misleading or distorted results comparing with some other alternatively defined ones.
}

\begin{document}
\maketitle
\flushbottom

\section{Introduction}\label{sec:intro}
CP violation (CPV) plays a crucial role in the Standard Model (SM)  and cosmology. In the SM, CPV originates from the complex elements of the Cabibbo-Kobayashi-Maskawa (CKM) matrix, which are associated with quark-level transition amplitudes ~\cite{Cabibbo:1963yz,Kobayashi:1973fv}. 
The experimental observations of CPV have been firmly established in the decays of $K$, $B$, and $D$ mesons ~\cite{Christenson:1964fg,KTeV:1999kad,BaBar:2001pki,Belle:2001zzw,Belle:2010xyn,BaBar:2010hvw,LHCb:2013syl,LHCb:2019hro} to date. Recently, the first observation of CPV in baryon decays was reported ~\cite{LHCb:2025ray}, providing valuable insights into the matter-antimatter asymmetry puzzle in the Universe.

CPV occurs via different mechanisms and can be expressed by different CPV observables in the decay
processes of hadrons. Typically, CP Asymmetries (CPAs) are induced by the interference of decay amplitudes with different weak phases. The most extensively studied type of CPV is the direct CPV which is characterized by the difference between the branching ratios of a pair of CP-conjugate decay processes. 
However, the interference terms containing CPV sometimes do not show up in this type of  CPV.  For example, some of the interference terms become zero after the differential decay widths are integrated over the whole phase space, rather, they show up non-trivially in the angular distributions, such as the triple product asymmetries of the momenta and/or spins of the involved particles \cite{Donoghue:1987wu,Valencia:1988it,Kayser:1989vw,Dunietz:1990cj,Bensalem:2000hq,Bensalem:2002pz,Bensalem:2002ys,Durieux:2015zwa,Gronau:2015gha,Shi:2019vus,Wang:2022fih}. Consequently, it is necessary to study the angular distributions in the case of direct CPV more extensively.

There are also some other typical examples of CPAs corresponding to the angular distributions. The first one corresponds to the Lee-Yang decay parameters
in the baryon decay $\Lambda\to p\pi^-$  \cite{Lee:1957qs}, where the CPV observable is conventionally defined as  \cite{Donoghue:1986hh}
\begin{equation}\label{eq:Acpalpha}
 A_{CP,\alpha}(\Lambda\to p\pi^-)\equiv \frac{\alpha-\overline{\alpha}}{\alpha+\overline{\alpha}}. \end{equation} 
The second example is the partial-wave CPAs (PWCPAs), which are introduced in multi-body decays of heavy hadrons \cite{Zhang:2021fdd}. 
Take a three-body decay as an example.
One first perform the Legendre expansion
to the decay amplitude squared (see Eq. (\ref{eq:Msqured}) below),
then the PWCPAs are conventionally defined as
\begin{equation} \label{eq:Acpconv}
A_{CP,l}^{\text{conv}}=\frac{w_l-\bar{w}_l}{w_l+\bar{w}_l},\\
\end{equation}
where $w_l$ and $\overline{w}_l$ are the weights in the Legendre expansion of the decay amplitude squared for a pair of CP-conjugate processes, respectively.

The definition of CPAs in the above examples are clearly not normalized, since neither the decay parameters such as $\alpha$ nor the weights in the Legendre expansions $w_l$ are positive definite.
The allowed values of $A_{CP,\alpha}$ and $A_{CP,l}^\mathrm{conv}$  are in principle in the range $(-\infty,+\infty)$.
The normalization problem for CPAs corresponding to the decay parameters of the hyperon decays is not serious since the CPAs in this case were measured to be small \cite{BESIII:2018cnd,BESIII:2022qax}.
Furthermore, this problem 
can be overcome by redefining the CPA observable as  \(A_{CP,\alpha}=\frac{1}{2}(\alpha+\overline{\alpha}),\) since $\alpha$ and $\bar{\alpha}$ are normalized by definition.
However, the PWCPAs defined in Eq. (\ref{eq:Acpconv}) cannot be solved in this way because  $w_l$ and $\overline{w}_l$ are not normalized.
There are also alternative definitions for these CPAs. 
For example, the PWCPAs can also be defined as \cite{Zhang:2025mne}
\begin{equation}
\label{eq:Acp0}
\mathring{A}_{CP,l}=\frac{w_l-\overline{w}_l}{w_0+\overline{w}_0}. 
\end{equation}
However, such definitions cannot solve the normalization problem either.
One of the motivation of the present paper is to solve such a normalization problem.

This paper is organized as follows. In Sec. \ref{sec:formalism}, we present an alternative definition of PWCPAs and discuss their normalization.
In Sec. \ref{sec:B23pi},  we apply our framework to the decays $B^\pm\rightarrow\pi^\pm\pi^+\pi^-$. We briefly give the summary in Sec. \ref{sec:summary}.

\section{The partial wave CP asymmetries and their normalization}\label{sec:formalism}

For a three-body decay process $M\rightarrow M_1M_2M_3$, the decay amplitude, $\mathcal{M}=\mathcal{M}(s_{12},s_{23})$, generally depends on the invariant masses of $M_1M_2$ and $M_2M_3$, denoted by $s_{12}$ and $s_{23}$, respectively. The invariant mass squared $s_{12}$ can be expressed as $s_{12}=\Delta_{12}\cos\theta+\Sigma_{12}$, where $\theta$ is the helicity angle which is defined as the angle between the momenta of $M_1$ and $M$ in the centre-of-mass frame of the $M_1M_2$ system, $\Delta_{12}=(s_{12,\text{max}}-s_{12,\text{min}})/2$, and $\Sigma_{12}=(s_{12,\text{max}}+s_{12,\text{min}})/2$, with $s_{12,\text{max}}$ and $s_{12,\text{min}}$ respectively being the kinematically allowed maximum and minimum values of $s_{12}$.
The kinematical parameter $s_{23}$ is irrelevant to our discussion here, so we simply ignore it in the following discussion.

The decay amplitude squared can be expanded by the Legendre polynomials as 
\begin{equation}\label{eq:Msqured}
\left|\mathcal{M}\right|^2=\sum_{l}w_l P_l(c_\theta),  
\end{equation}
where $P_l$ is the $l$-th Legendre polynomial, $w_l$ is, as is introduced in the previous section, the corresponding weight in the expansion.
We define the PWCPAs as 
\begin{equation}
\label{eq:ACPl}
A_{CP,l}=\eta_l\mathring{A}_{CP,l} =\eta_l\frac{w_l-\overline{w}_l}{w_0+\overline{w}_0}=\frac{\eta_l(2l+1)\int \left[\left|\mathcal{M}\right|^2-\left|\overline{\mathcal{M}}\right|^2\right] P_l(c_\theta) \tilde{d}c_\theta}{\int  \left[\left|\mathcal{M}\right|^2+\left|\overline{\mathcal{M}}\right|^2\right] \tilde{d}c_\theta},  
\end{equation}
where $\overline{\mathcal{M}}$ is the decay amplitude for the CP-conjugate process,  $\tilde{d}c_\theta\equiv\Delta_{12} dc_\theta$, and $\eta_l$ can be viewed as a set of normalization factors. 
The motivation for  introducing the factors  $\eta_l$ is to ensure that the PWCPAs are comparable with each other, as well as with other CPA observables, so that one can compare these CPA observables in a reasonable way, and determine which CPAs are expected to  be observed more easily in experiments.
The guaidlines for determining these factors $\eta_l$ will be presented in the following. 

In order to obtain a proper form of the factors $\eta_l$, let us first introduce a set of experiment-friendly CPA observables, 
\begin{equation}
\label{eq:hatACPl}
\hat{A}_{CP,l}\equiv \frac{\int \left[\left|\mathcal{M}\right|^2-\left|\overline{\mathcal{M}}\right|^2\right]  \text{sgn}(P_l(c_\theta)) \tilde{d}c_\theta}{\int \left[\left|\mathcal{M}\right|^2+\left|\overline{\mathcal{M}}\right|^2\right] \tilde{d}c_\theta}=\frac{\sum_i (N_i-\overline{N}_i)\text{sgn}_{l,i}}{N+\overline{N}},   
\end{equation} 
where in the second equality we have divided the interval of $c_{\theta}$ into small bins (denoted by $i$ in the subscript), and express $\hat{A}_{CP,l}$ by $N_i$ and $\overline{N}_i$,  the event yields of the CP-conjugate processes in each bin, respectively, and $N$ and $\overline{N}$,  the corresponding total ones. The sign functions in Eq. (\ref{eq:hatACPl}) are defined as $\text{sgn}(x)=\pm1$, if $x \gtrless 0$,  and $\text{sgn}_{l,i}$ is the abbreviation for $\text{sgn}(P_l(c_{\theta_i}))$. It can be easily seen that 1) $\hat{A}_{CP,l}$'s satisfy the normalization condition $-1\leq \hat{A}_{CP,l}\leq +1$, 2) for $l=0$, $\hat{A}_{CP,l}$ is just the direct CPA, and 3) the statistical errors of $\hat{A}_{CP,l}$'s are approximately the same, i.e., \(\sigma_{\hat{A}_{CP,l}}\approx\hat{\sigma}\), where the $\hat{\sigma}$ is the statistical error which takes the familiar form $1/\sqrt{N+\overline{N}}$ in the absence of the background.
In principle, both sets of observables $\{{A}_{CP,l}\}$ and $\{\hat{A}_{CP,l}\}$ are equivalent.
Moreover, the latter can be expressed in terms of the former as
\begin{equation}\label{eq:hatAcp1}
\hat{A}_{CP,l}=\sum_k\Omega_{lk}{A}_{CP,k}, \end{equation}where $\Omega$ is the matrix connecting these two sets of CPA observables, the elements of which can be easily shown to take the form $\Omega_{lk}=\frac{\omega_{lk}}{\eta_k}$, where
\begin{equation}\label{eq:omegalk}
\omega_{lk} =\frac{1}{2}\int_{-1}^{+1} \text{sgn}\left(P_l(c_\theta)\right)P_k(c_\theta)dc_\theta.
\end{equation}
The matrix elements $\omega_{lk}$ have the following properties: 1) they take non-zero values only when $l$ and $k$ are both even or odd; 2) the diagonal elements are much larger than the off-diagonal ones in the same rows or columns.

As a first approximation, we demand that the diagonal elements of $\Omega$ equal 1, so we have
\begin{equation}\label{eq:hatAcp2}
\hat{A}_{CP,l}={A}_{CP,l} + \sum_{k\neq l}\Omega_{lk}{A}_{CP,k}.    
\end{equation}
The advantages for this choice is clear, 
the elements of $\Omega_{lk}$ are quite smaller than 1 for $k\neq l$, hence one can see transparently that $\hat{A}_{CP,l}$ will get the most important contribution from ${A}_{CP,l}$.
Since $\hat{A}_{CP,l}$ are clearly normalized, we conclude that ${A}_{CP,l}$ are now {\it roughly} normalized.
The choice of the diagonal elements of $\Omega$ implies that the factors $\eta_l$ are chosen as
\begin{equation}\label{eq:etal0}
\tilde{\eta}_l=\omega_{ll}=\frac{1}{2}\int_{-1}^{+1} \text{sgn}\left(P_l(c_\theta)\right)P_l(c_\theta)dc_\theta=\frac{1}{2}\int_{-1}^{+1} \left|P_l(c_\theta)\right|dc_\theta,
\end{equation}
where the tilde above $\eta_l$ indicates that $\tilde{\eta}_l$ should be viewed as the first approximation for the factors $\eta_l$.

Interestingly, besides the fact that ${A}_{CP,l}$'s are roughly normalized, one can also see that they share roughly the same statistical errors with $\hat{A}_{CP,l}$.
This loose correlation inspires us to pursue more practical and reasonable estimations of $\eta_l$.
To this end, we 
assume that the Legendre expansion is truncated at some finite number $L$, so that the matrix $\Omega$ and $\omega$ become dimension $L+1$ ones, which will be denoted as $\Omega_L$ and $\omega_L$, respectively.
It can be shown that the matrices $\Omega_L$ and $\omega_L$ are reversible. 
This allows us to express $A_{CP,l}$ inversely in terms of $\hat{A}_{CP,k}$:
\begin{equation}\label{eq:ACP2hatACP}
A_{CP,l}=\sum_{k=0}^{L}(\Omega_L^{-1})_{lk}\hat{A}_{CP,k}=\eta_l \sum_{k=0}^{L}(\omega_L^{-1})_{lk}\hat{A}_{CP,k},
\end{equation}
where $\Omega_L^{-1}$ and $\omega_L^{-1}$ are the inverse matrices of $\Omega_L$ and $\omega_L$, respectively, and the relation $(\Omega_L^{-1})_{lk}=\eta_l (\omega_L^{-1})_{lk}$ has been used.
The errors of $A_{CP,l}$ can be expressed as
\begin{equation}\label{eq:sigma}
\sigma_l^2= \left[\Omega_L^{-1}\hat{\rho}\left(\Omega_L^{-1}\right)^T\right]_{ll}\hat{\sigma}^2=\eta_l^2\left[\omega_L^{-1}\hat{\rho}\left(\omega_L^{-1}\right)^T\right]_{ll}\hat{\sigma}^2,
\end{equation}
where $\hat{\rho}_{kk'}$ are the correlation coefficients between $\hat{A}_{CP,k}$ and $\hat{A}_{CP,k'}$, which, as will be shown below, can be well estimated as
\begin{equation}\label{eq:CorMatrix}
\hat{\rho}_{kk'}\approx \frac{1}{2}\int_{-1}^{+1}\text{sgn}(P_k(c_\theta))\text{sgn}(P_{k'}(c_\theta))dc_\theta .
\end{equation}
Instead of the requirement of normalization, it would be practically more useful to choose the $\eta_l$'s in such a way  that  {\it the statistical error of $A_{CP,l}$ also equals to $\hat{\sigma}$, i.e., $\sigma_l=\hat{\sigma}$.}
This means that $\eta_l$ should be chosen as
\begin{equation}\label{eq:etalL}
\eta_l^{(L)}=\frac{1}{\sqrt{\left[\omega_L^{-1}\hat{\rho}(\omega_L^{-1})^T\right]_{ll}}}=\frac{1}{\sqrt{\sum_{k,k'=0}^{L}(\omega_L^{-1})_{lk}\omega_L^{-1})_{lk'}\hat{\rho}_{kk'}}},~~l=0,1,\cdots,L,
\end{equation}
as is indicated by Eq. (\ref{eq:sigma}),
where the subscript ``$L$'' indicates that the results are truncation-dependent.
This is a more accurate estimation for the factors $\eta_l$ comparing with Eq. (\ref{eq:etal0}), in the sense that the statistical error of $A_{CP,l}$ is closer to that of $\hat{A}_{CP,l}$.

In the following we present the reasons for the expression of the correlation coefficients $\hat{\rho}_{kk'}$ in Eq. (\ref{eq:CorMatrix}).
The correlation matrix is defined according to
\[
\hat{\sigma}^2\hat{\rho}_{lk}=\sum_{i}\left(\frac{\partial \hat{A}_{CP,k}}{\partial N_i}\frac{\partial \hat{A}_{CP,l}}{\partial N_i}\sigma_{N_i}^2+\frac{\partial \hat{A}_{CP,l}}{\partial \overline{N}_i} \frac{\partial\hat{A}_{CP,k} }{\partial \overline{N}_i}\sigma_{\overline{N}_i}^2\right).
\]
Hence $\hat{\rho}_{lk}$ takes the form
\begin{align*}   \hat{\rho}_{lk}=\frac{1}{(N+\overline{N})^2}\sum_i\Bigg\{ \frac{\sigma_{N_i}^2+\sigma_{\overline{N}_i}^2}{\hat{\sigma}^2}\left[\text{sgn}_{l,i}\text{sgn}_{k,i} + \frac{\sum_{j}(N_j-\overline{N}_j)\text{sgn}_{l,j} \sum_{j'}(N_{j'}-\overline{N}_{j'})\text{sgn}_{k,j'}}{(N+\overline{N})^2}\right]\\
-\frac{\sigma_{N_i}^2-\sigma_{\overline{N}_i}^2}{\hat{\sigma}^2}\left[
\frac{\text{sgn}_{l,i}\sum_j(N_j-\overline{N}_j)\text{sgn}_{k,j} + \text{sgn}_{k,i}\sum_j(N_j-\overline{N}_j)\text{sgn}_{l,j}}{N+\overline{N}}\right]\Bigg\}. 
\end{align*}
In the limit of CP symmetry, we obtain \footnote{Note that this equation also works when the contributions of the background are included in the statistical errors. Typically, the background may rescale the statistical errors, for example, the error $\hat{\sigma}$ would be rescaled to $\lambda/\sqrt{N+\overline{N}}$, where $\lambda$ is the rescale factor. Meanwhile, the errors $\sigma_{N_i}$ and $\sigma_{\overline{N}_i}$ will also be rescaled to $\lambda\sqrt{N_i}$ and $\lambda\sqrt{\overline{N}_i}$. Consequently, because of the cancellation of the rescale factors, one obtain $\frac{\sigma_{N_i}^2+\sigma_{\overline{N}_i}^2}{\hat{\sigma}^2}=(N_i+\overline{N}_i)(N+\overline{N})$.}
\[
\hat{\rho}_{lk}\approx \frac{1}{(N+\overline{N})^2} \sum_i \left(\frac{\sigma_{N_i}^2+\sigma_{\overline{N}_i}^2}{\hat{\sigma}^2}\right) \text{sgn}_{l,i}\text{sgn}_{k,i}=\frac{\sum_i\left({N_i}+{\overline{N}_i}\right) \text{sgn}_{l,i}\text{sgn}_{k,i}}{N+\overline{N}}.  
\]
This is clearly a very good approximation if the CPAs are not quite large.
Not only that, at the end of this section we will show that there is no need for the assumption of CP symmetry at all.
At first sight, $\hat{\rho}$ depends on the event distributions, or the decay amplitudes, which is a disadvantage since it means that $\rho$ is not universal.
However, as one will see that this dependence is quite mild and can be neglected safely.
First of all, one can see that the diagonal elements of $\hat{\rho}$ equal 1, hence are independent of the decay amplitudes and are universal.
Secondly, although the off-diagonal elements depend on the decay amplitudes, 
they are much smaller than the diagonal ones,
which means that we can safely assume  plain distributions of the event yields for estimation, so that a universal estimation of the matrix elements of $\hat{\rho}$ are obtained in Eq. (\ref{eq:CorMatrix}).
Based on the same argument, it is expected that the factors $\eta^{(L)}_l$ obtained in Eq. (\ref{eq:etalL}) should be close to the uncorrelated limit, i.e., $\hat{\rho}_{kk'}=\delta_{kk'}$.

Taking $L=4$ as an example. 
One first calculates the numerical values of the matrix $\omega$ and $\hat{\rho}$ respectively according to Eqs. (\ref{eq:omegalk}) and (\ref{eq:CorMatrix}), which read
\[\omega= 
\begin{pmatrix}
1    & 0   & 0    &0 &0\\
0    & 0.5 & 0    & -0.1250&0\\
-0.1547& 0   & 0.3849 & 0 &-0.0642\\
0    &-0.1 & 0    & 0.3250 &0\\
-0.0400&0    &-0.0768&0     & 0.2866\\
\end{pmatrix}, \]
and 
\[\hat{\rho}= 
\begin{pmatrix}
1    & 0   & -0.1547    &0 &-0.0423\\
0    & 1 & 0    & -0.5492  & 0\\
-0.1547& 0   & 1 & 0 &-0.2475\\
0    &-0.5492 & 0    & 1 &0\\
-0.0423 &0    &-0.2475 & 0     &1 \\
\end{pmatrix},\] 
respectively \footnote{One can see from $\hat{\rho}$ that the largest correlation occurs between $\hat{A}_{CP,1}$ and $\hat{A}_{CP,3}$.}.
Then substituting these results into Eq. (\ref{eq:etalL}), one obtains 
\begin{equation}\label{eq:etal4}
\eta^{(4)}_0=1,~\eta^{(4)}_1=0.5418,~\eta^{(4)}_2=0.3847,~\eta^{(4)}_3=0.3312, ~\eta^{(4)}_4=0.2829.
\end{equation}
For comparison, the factors $\eta_l$ for the uncorrelated limit read
\begin{equation}
\label{eq:etalUnco}
\left.\eta^{(4)}_0\right|_{\hat{\rho}=\mathbb{1}}=1,~\left.\eta^{(4)}_1\right|_{\hat{\rho}=\mathbb{1}}=0.4308,~\left.\eta^{(4)}_2\right|_{\hat{\rho}=\mathbb{1}}=0.3540,~\left.\eta^{(4)}_3\right|_{\hat{\rho}=\mathbb{1}}=0.2942, ~\left.\eta^{(4)}_4\right|_{\hat{\rho}=\mathbb{1}}=0.2674,
\end{equation}
and the first few $\tilde{\eta}_l$'s based on the rough estimation in Eq. (\ref{eq:etal0}) take the following values: 
\begin{equation}
\label{eq:etal0num}
\tilde{\eta}_0=1, ~~\tilde{\eta}_1=0.5,~~\tilde{\eta}_2=0.3849,~~\tilde{\eta}_3=0.3250,~~\tilde{\eta}_4=0.2866.\end{equation}
The comparison of Eqs. (\ref{eq:etal4}), (\ref{eq:etalUnco}), and (\ref{eq:etal0num}) shows that 1) the contributions to the factors $\eta_l$ from the correlation between the statistical errors of different CPA observables $A_{CP,l}$ are small, 2)
the rough estimation of the $\eta_l$ based on Eq. (\ref{eq:etal0}), i.e., $\tilde{\eta}_l$, works pretty well.
The results of $\eta^{(4)}_l$ in Eq. (\ref{eq:etal4}), together with the numerical values of $\tilde{\eta}_l$ in Eq. (\ref{eq:etal0num}), will be adopted in the application in the next section.
More numerical values of $\eta^{(L)}_l$, $\left.\eta^{(L)}_l\right|_{\hat{\rho}=\mathbb{1}}$, and $\tilde{\eta}$ are presented in the Table \ref{tab:etalL}.
As one can clearly see from this table, the truncation-dependence of $\eta^{(L)}_l$ and $\left.\eta^{(L)}_l\right|_{\hat{\rho}=\mathbb{1}}$ is small.

\begin{table*}
	\centering
\begin{tabular}
{|c|c||m{.9cm}|m{.9cm}|m{.9cm}|p{.9cm}|m{.9cm}|m{.9cm}|m{.9cm}|p{.9cm}|m{.9cm}|m{.9cm}|m{.9cm}|}
\hline
&\diagbox[width=1cm,height=1cm]{$L$}{$l$} & 0 & 1 & 2 & 3 & 4 & 5 & 6 & 7 & 8\\ \hline\hline
\multirow{7}{*}{$\eta_{l}^{(L)}$}&0 & 1 & & & & & & & &\\ \cline{2-11}
&1 & 1 & 0.5 & & & & & & &\\ \cline{2-11}
&2 & 1 & 0.5 & 0.3896& & & & & &\\ \cline{2-11}
&3 & 1 & 0.5418 & 0.3896 & 0.3312 & & & & & \\ \cline{2-11}
&4 & 1 & 0.5418 & 0.3847 & 0.3312 & 0.2829 & & & &\\ \cline{2-11}
&5 & 1 & 0.5552 & 0.3847 & 0.3332 & 0.2829 & 0.2615 & & &\\ \cline{2-11}
&6 & 1 & 0.5552 & 0.3899 & 0.3332 & 0.2807 & 0.2615 & 0.2345 & &\\ \cline{2-11}
&7 & 1 & 0.5656 & 0.3899 & 0.3312 & 0.2807 & 0.2630 & 0.2345 & 0.2205 &\\ \cline{2-11}
&8 & 1 & 0.5656 & 0.3967 & 0.3312 & 0.2793 & 0.2630 & 0.2322 & 0.2205 & 0.2064\\
\cline{2-11}
\hline\hline
\multirow{7}{*}{$\left.\eta_{l}^{(L)}\!\right|_{\hat{\rho}=\mathbb{1}}$} & 0 & 1 & & & & & & & & \\ \cline{2-11}
&1 & 1 & 0.5 & & & & & & &\\ \cline{2-11}
&2 & 1 & 0.5 & 0.3804& & & & & &\\ \cline{2-11}
&3 & 1 & 0.4308 & 0.3804 & 0.2942 & & & &&\\ \cline{2-11}
&4 & 1 & 0.4308 & 0.3540 & 0.2942 & 0.2674 & & & &\\ \cline{2-11}
&5 & 1 & 0.4383 & 0.3540 & 0.2787 & 0.2674 & 0.2441 & & &\\ \cline{2-11}
&6 & 1 & 0.4383 & 0.3200 & 0.2787 & 0.2423 & 0.2441 & 0.2131 & & \\
\cline{2-11}
&7 & 1 & 0.4256 & 0.3200 & 0.2668 & 0.2423 & 0.2232 & 0.2131 & 0.1998 & \\ \cline{2-11}
&8 & 1 & 0.4256 & 0.3252 & 0.2668 & 0.2378 & 0.2232 & 0.1986 & 0.1998 & 0.1915\\
\hline \hline
$\tilde{\eta}_l$& & 1 & 0.5 & 0.3849 & 0.3250 & 0.2866 & 0.2592 & 0.2385 & 0.2220 & 0.2086\\ \hline
\end{tabular}
\caption{The numerical values of $\eta_{l}^{(L)}$, $\left.\eta_{l}^{(L)}\right|_{\hat{\rho}=\mathbb{1}}$ (for truncation $L\leq 8$), and $\tilde{\eta}_l$ (for $l\leq 8$).}\label{tab:etalL}
\end{table*}

At the end of this section, we want to point out that 
the factors $\eta_l$ in fact reflect the bound of the ratio $w_l/w_0$.
Hence our mathematical conclusion is that $\left|w_l/w_0\right|$ is bounded roughly by $1/\eta_l$, with $\eta_l$ (at least approximately) taking the forms in Eq. (\ref{eq:etalL}), or even more simply, in Eq. (\ref{eq:etal0}).
This means that
the introduction and the determination of the factors $\eta_l$ can be done without the introduction of PWCPAs.
To see this, we first introduce
the PWCPAs alternatively as $\tilde{A}_{CP,l}=(A_l-\overline{A}_l)/2$, where $A_l=\eta_l w_l/w_0$ and $\overline{A}_l=\eta_l\overline{w}_l/\overline{w}_0$ \cite{Zhang:2022emj}
\footnote{Clearly, $\tilde{A}_{CP,l}$ and ${A}_{CP,l}$ are equivalent. However, the physical meaning of ${A}_l$ and $\overline{A}_l$ for even $l$ seems not transparent. This is the reason why we did not perform the analysis with $\tilde{A}_{CP,l}$, as well as ${A}_l$ and $\overline{A}_l$, in the main text.}. 
Let us focus on $A_l$. 
One can introduce a set of experiment-friendly observables $\hat{A}_{l}\equiv\frac{\int \left[\left|\mathcal{M}\right|^2\right]  \text{sgn}(P_l(c_\theta)) \tilde{d}c_\theta}{\int \left[\left|\mathcal{M}\right|^2\right] \tilde{d}c_\theta}=\frac{ \sum_iN_i\text{sgn}_{l,i}}{N}$, and perform almost exactly the same procedure, in this way one can obtain exactly the same expressions for $\tilde{\eta}_l$ and  $\eta^{(L)}_l$. Moreover, the same procedure clearly also applies to $\overline{A}_l$, so that both $\tilde{\eta}_l$ and $\eta^{(L)}_l$ should be the same for the pair of CP-conjugate processes, regardless how large the CPAs are.
Clearly, neither $\tilde{\eta}_l$ nor $\eta_l^{(L)}$ can ensure that $A_l$ or $\overline{A}_l$, as well as $\tilde{A}_{CP,l}$ or $A_{CP,l}$, are normalized, but with our choices of $\tilde{\eta}_l$ or $\eta_l^{(L)}$, they must be not far away from normalization. Consequently, we will call them quasi-normalized ones.
For example, with our choices of $\eta_{l}$ in Eqs. (\ref{eq:etal0}) or (\ref{eq:etalL}), $A_{CP,l}$ will be called as quasi-normalized PWCPAs.

\section{Application to $B^\pm\rightarrow\pi^\pm\pi^+\pi^-$ decays}\label{sec:B23pi} 
As an application, we present the analysis of the PWCPAs for the decays $B^\pm\rightarrow\pi^\pm\pi^+\pi^-$ in this section. We  perform this analysis based on the data of the LHCb collaboration in Ref. \cite{LHCb:2019sus},
focusing on the phase space where the invariant mass  of the $\pi^+\pi^-$ pair $m_{\pi\pi,\mathrm{low}}$ is around the vicinity of the $\rho^0(1450)$ resonance. 
There are several resonances that may contribute to this decay, including $f_0(1370)$, $f_0(1500)$, $f_2(1270)$, $f_2(1430)$, $f'_2(1525)$, etc.. 
After various fitting scenarios were attempted, we found
that the best fit to the LHCb data of Fig. 12 in Ref. \cite{LHCb:2019sus}  for $m_{\pi\pi,\mathrm{low}}$  ranging between 1.262 and 1.676 GeV can be obtained with the resonances $\rho^0(1450)$, $f_0(1500)$, and $f_2(1270)$ with the quantum numbers $J^{PC}$ being $1^{--}$, $0^{++}$, and $2^{++}$, respectively. 
This means that the Legendre expansion of the decay amplitude squared should be truncated at $l=4$, which reads 
\begin{equation} \label{eq:MtotL4}
|\mathcal{M}^{\pm}|^2=\sum_{l=0}^4w^\pm_lP_l(c_\theta),\\
\end{equation}
where ``$\pm$'' in the superscript indicates the electric charge of the $B$ meson.
The weights of the expansion $w^{\pm}_l$ are found to take the forms
\begin{equation} \label{eq:wRl}
\begin{split}
w^\pm_0&=\Bigg(|\mathcal{M}^{\pm}_{f_0}|^2+\frac{1}{5}|\mathcal{M}^{\pm}_{f_2}|^2+\frac{1}{3}|\mathcal{M}^{\pm}_{\rho^0}|^2\Bigg),\quad
w^\pm_1=\Bigg(2\mathrm{Re}[\mathcal{M}^{\pm}_{\rho^0}{\mathcal{M}^{\pm}_{f_0}}^*]+\frac{4}{5}\mathrm{Re}[\mathcal{M}^{\pm}_{\rho^0}{\mathcal{M}^{\pm}_{f_2}}^*]\Bigg),\\
w^\pm_2&=\Bigg(\frac{2}{3}|\mathcal{M}^{\pm}_{\rho^0}|^2+\frac{2}{7}|\mathcal{M}^{\pm}_{f_2}|^2+2\mathrm{Re}[\mathcal{M}^{\pm}_{f_0}{\mathcal{M}^{\pm}_{f_2}}^*]\Bigg),\quad w^\pm_3=\frac{6}{5}\mathrm{Re}[\mathcal{M}^{\pm}_{\rho^0}{\mathcal{M}^{\pm}_{f_2}}^*],\\
w^\pm_4&=\frac{18}{35}|\mathcal{M}^{\pm}_{f_2}|^2,\\
\end{split}
\end{equation}
where $\mathcal{M}^{\pm}_{R}$ represent the decay amplitudes of the cascade decays $B^\pm \rightarrow R(\to\pi^+\pi^-)\pi^\pm$, with $R=f_0$, $\rho^0$, and $f_2$ representing the resonances $f_0(1500)$,   $\rho^0(1450)$ and $f_2(1270)$, respectively.
The decay amplitudes are parameterized as 
\begin{equation}\label{eq:MR}
\begin{split}
\mathcal{M}^{\pm}_{R}&= c^{\pm}_RF_R^{BW}e^{i\delta^{\pm}_R}P_l(c_\theta),\\
\end{split}
\end{equation}
 where $F_R^{BW}$ is the the Breit-Wigner line-shape of resonance $R$, 
 $c^{\pm}_R$ and $\delta^{\pm}_R$ are the corresponding
amplitude
and the relative phase, respectively.

 \begin{figure}[H]
\centering
\subfigure[]{
\includegraphics[width=.48\textwidth]{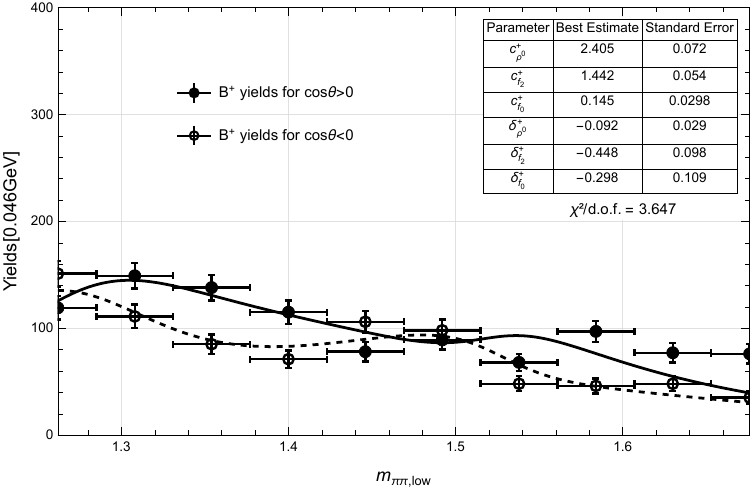}}
\subfigure[]{
\includegraphics[width=.48\textwidth]{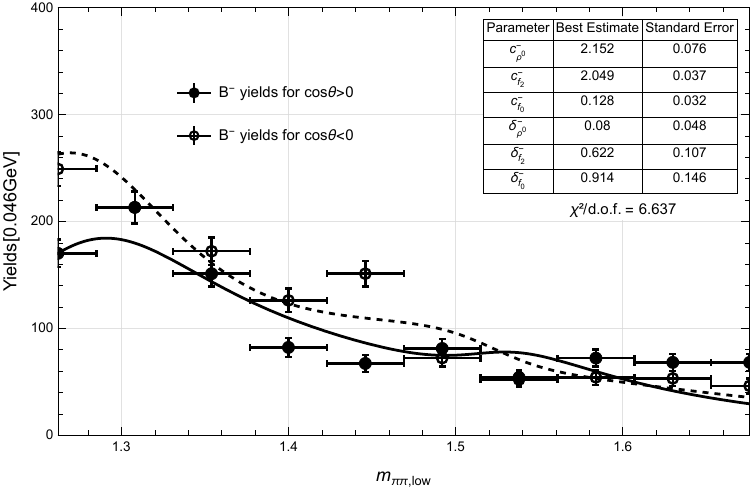}}
\caption{The fitting results of the parameters (shown in the upper right corner of each figure) from the event yields of $B^+\to\pi^+\pi^-\pi^+$ (a) and $B^-\to\pi^+\pi^-\pi^-$ (b) for $\cos\theta>0$ and $\cos\theta<0$. }
\label{fig:Fitresult}
\end{figure}

\begin{figure}[H]
\centering
\includegraphics[width=.9\textwidth]{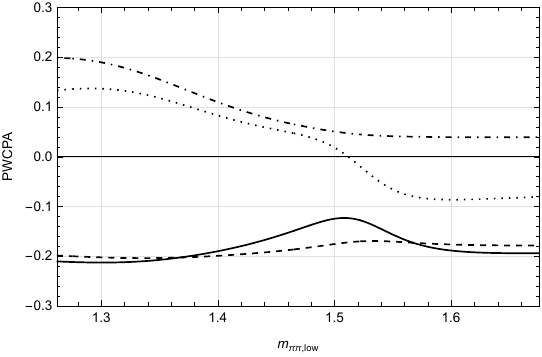}
\caption{Predictions of the quasi-normalized PWCPAs $\left.A_{CP,l}\right|_{\eta_l=\eta_l^{(4)}}$ for $B^\pm\to \pi^+\pi^-\pi^\pm$ as a function of the invariant mass $m_{\pi\pi,\mathrm{low}}$. The solid, dotted, dashed, and dash-dotted lines correspond to $l=1$, 2, 3, and 4, respectively. }
\label{fig:ACPl}
\end{figure}

\begin{figure}[H]
\centering
\subfigure[]{
\includegraphics[width=.48\textwidth]{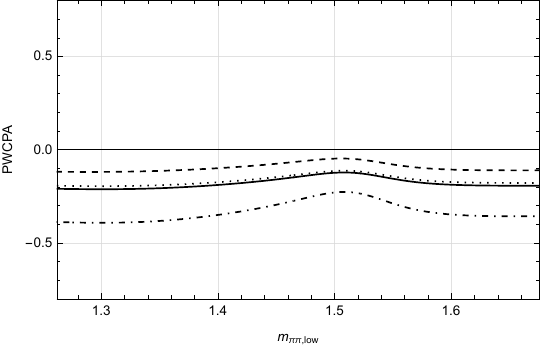}\label{fig:ACPlcom_a}}
\subfigure[]{
\includegraphics[width=.48\textwidth]{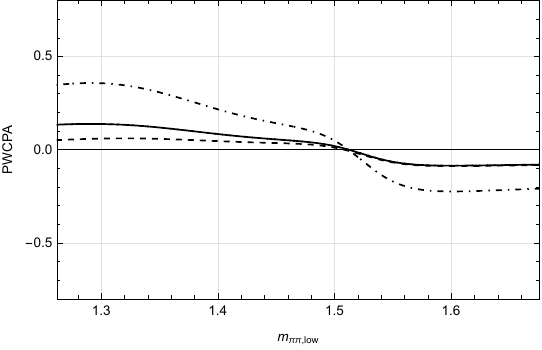}\label{fig:ACPlcom_b}}
\subfigure[]{
\includegraphics[width=.48\textwidth]{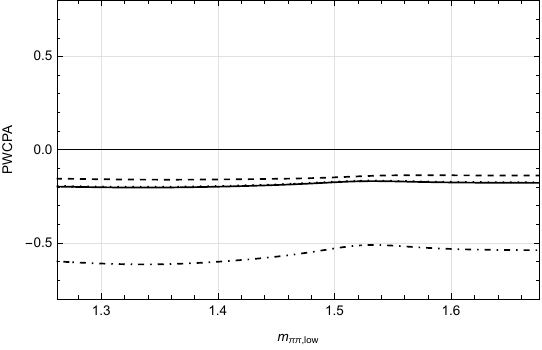}\label{fig:ACPlcom_c}}
\subfigure[]{
\includegraphics[width=.48\textwidth]{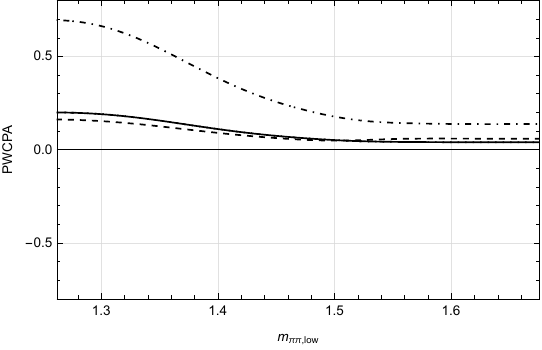}\label{fig:ACPlcom_d}}
\caption{Comparisons of various definitions of PWCPAs, where the solid, dotted,  dashed , and dot-dashed lines correspond to $\left.A_{CP,l}\right|_{\eta_l=\eta_l^{(4)}}$, $\left.A_{CP,l}\right|_{\eta_l=\tilde{\eta}_l}$, $\hat{A}_{CP,l}$, and  $\mathring{A}_{CP,l}$, respectively. (a), (b), (c), and (d) correspond to $l=1$, 2, 3, and 4, respectively. Note that $\left.A_{CP,l}\right|_{\eta_l=\eta_l^{(4)}}$ and  $\left.A_{CP,l}\right|_{\eta_l=\tilde{\eta}_l}$ almost coincide with each other.}
\label{fig:ACPlcom}
\end{figure}

The fitting results are presented in Fig. \ref{fig:Fitresult}.
With the values of the parameters $c_R^\pm$ and $\delta_R^\pm$, we are able to make predictions for the quasi-normalized PWCPAs, which are defined in Eq. (\ref{eq:ACPl}), and the results with  $\eta_l$ taking the values $\eta_l^{(4)}$ are presented in Fig. \ref{fig:ACPl}.
The comparison of PWCPAs with different definitions are presented in Fig. \ref{fig:ACPlcom}.

One simple fact about the quasi-normalized PWCPAs from Fig. \ref{fig:ACPl} is that all the four PWCPAs, $A_{CP,l}$ (l=1, 2, 3, 4), range roughly around [-0.20, 0.20]. 
Note that this is a quite strong conclusion and depends crucially on the choice of the factors $\eta_l$. 
The four diagrams in Fig. \ref{fig:ACPlcom} provide strong evidences that the choices of the factors $\eta_l$ in Eqs. (\ref{eq:etal0}) and (\ref{eq:etalL}) lead to quite reasonable relative sizes of all the four quasi-normalized PWCPAs in Fig. \ref{fig:ACPl}.
These evidences, as can be seen 
from Fig. \ref{fig:ACPlcom}, include
1) $\left.A_{CP,l}\right|_{\eta_l=\eta_l^{(4)}}$ and $\left.A_{CP,l}\right|_{\eta_l=\tilde{\eta}_l}$ are very close to $\hat{A}_{CP,l}$;
2)
$\left.A_{CP,l}\right|_{\eta_l=\eta_l^{(4)}}$ and $\left.A_{CP,l}\right|_{\eta_l=\tilde{\eta}_l}$ are very close to each other; and
3) $\mathring{A}_{CP,l}$ are quite different form $\left.A_{CP,l}\right|_{\eta_l=\eta_l^{(4)}}$, $\left.A_{CP,l}\right|_{\eta_l=\tilde{\eta}_l}$, and $\hat{A}_{CP,l}$.

From Fig. \ref{fig:ACPlcom_a} one can also see that the largest difference between $\left.A_{CP,l}\right|_{\eta_l=\eta_l^{(4)}}$ ($\left.A_{CP,l}\right|_{\eta_l=\tilde{\eta}_l}$) and $\hat{A}_l$ occurs when $l=1$.
This is understandable.
According to Eq. (\ref{eq:hatAcp2}), we have $\hat{A}_{CP,1}=A_{CP1}+(\omega_{13}/\tilde{\eta}_{3}) A_{CP,3}=A_{CP1}+(-0.1250/0.3250) A_{CP,3}\approx A_{CP1}-0.385 A_{CP,3}$.
Since $A_{CP,3}$ takes the value of about 0.2, the deviation of $\hat{A}_{CP,1}$ from $A_{CP,1}$ is as large as about 0.08, which is just the difference between $\hat{A}_{CP,1}$ and $A_{CP,1}$ in Fig. \ref{fig:ACPlcom_a}.

\begin{figure}[H]
\centering
\includegraphics[width=.9\textwidth]{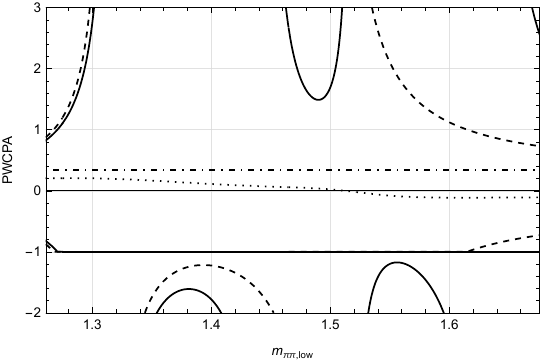}
\caption{The conventionally defined PWCPAs, $A_{CP,l}^{\text{conv}}$,  as a function of $m_{\pi\pi,\text{low}}$, where the solid, dotted, dashed and dash-dotted lines correspond to $l=1$, 2, 3, and 4, respectively. 
We also present the results for another alternative PWCPAs, which are defined as $\tilde{A}_{CP,l}^{\text{conv}}\equiv \frac{w_l-\bar{w}_l}{|w_l|+|\bar{w}_l|}$, with $l$ corresponding to the same line types as those for $A_{CP,l}^{\text{conv}}$.
For $l=1$ and 3, $\tilde{A}_{CP,l}^{\text{conv}}$ are
 exactly $-1$ in the main part of the region of $m_{\pi\pi,\text{low}}$; while for $l=2$ and 4, $\tilde{A}_{CP,l}^{\text{conv}}$ exactly coincides with  ${A}_{CP,l}^{\text{conv}}$.}
\label{fig:ACPlconv}
\end{figure}

For completeness, we also present the conventionally defined PWCPAs, $A_{CP,l}^{\mathrm{conv}}$, in Fig. \ref{fig:ACPlconv}.
Clearly, $A_{CP,l}^{\mathrm{conv}}$'s are not normalized.
Even worse, they are not even bounded sometime, as can been clearly seen from $A_{CP,1}^{\mathrm{conv}}$ and $A_{CP,3}^{\mathrm{conv}}$ in this figure.
Simply taking the absolute values of $w_l$ and $\overline{w}_l$ in the denominator of the definition of $A_{CP,1}^{\mathrm{conv}}$ does not solve this problem.
It only forces $A_{CP,l}^{\mathrm{conv}}$
to be equal to $\pm 1$ when $|A_{CP,l}^{\mathrm{conv}}|>1$. This is also clearly illustrated in Fig. \ref{fig:ACPlconv}, by the solid and dashed lines which are $-1$ in the main part of the $m_{\pi\pi,\mathrm{low}}$ range.
The comparison between Figs. \ref{fig:ACPl} and \ref{fig:ACPlconv} shows the large differences between the quasi-normalized and the conventionally defined PWCPAs.
These differences indicate clearly that it is necessary to study the PWCPAs with the quasi-normalized ones in order to avoid  misleading or distorted results.

\section{Summary}\label{sec:summary}
In this paper, we pointed out the problem of the normalization of the PWCPAs in multi-body decay processes of heavy hadrons.
To solve this problem, we redefined the PWCPAs according to Eq. (\ref{eq:ACPl}), which are called as the quasi-normalized ones in this paper.
To obtain the proper forms of the normalization factors $\eta_l$, the experiment-friendly CPV observables $\hat{A}_{CP,l}$'s, which are clearly normalized and have a proper form of the statistical errors, were introduced.
These properties of $\hat{A}_{CP,l}$, together with the correlation between $\hat{A}_{CP,l}$ and $A_{CP,l}$, allowed us to derive reasonable estimations of the factors $\eta_l$ in Eqs. (\ref{eq:etal0}) and (\ref{eq:etalL}), where the former was obtained based on rough estimation, while the latter was obtained by requiring that the statistical errors of ${A}_{CP,l}$ and $\hat{A}_{CP,l}$ be approximately the same. 
The numerical values of the factors $\eta_l$ were also presented in Table \ref{tab:etalL} for future reference. Interestingly and reasonably, the numerical values of $\eta_l$ obtained based on Eqs. (\ref{eq:etal0}) and (\ref{eq:etalL}) are quite close to each other.

As an illustration, we presented an analysis of  the quasi-normalized PWCPAs in Eq. (\ref{eq:ACPl}) for the decay processes $B^\pm\to \pi^+\pi^-\pi^\pm$.
With the amplitudes extracted from the data of the LHCb collaboration in Ref. \cite{LHCb:2019sus}, we were able to obtain the quasi-normalized PWCPAs in the region around the intermediate resonance $\rho^0(1450)$.
The results illustrated in Figs. \ref{fig:ACPl}, \ref{fig:ACPlcom}, and \ref{fig:ACPlconv} show clearly that the quasi-normalized PWCPAs indeed can overcome the deficiencies of some other alternatively defined PWCPAs.




\acknowledgments
We thank Prof. Bing-Song Zou (Tsinghua University),  Prof. Hsing-nan Li (Institute of Physics, Academia Sinica), Prof. Haiyong Wang (School of Mathematics and Statistics, Huazhong University of Science and Technology), Prof. Qin Qin (Huazhong University of Science and Technology), Prof. Fu-Sheng Yu (Lanzhou University) and Dr. Tian-Liang Feng (Lanzhou University) for valuable discussions.
This work was supported by National Natural Science Foundation of China under Grants Nos. 12475096, 12275024, 12405115 and 12105149,  Scientific Research Fund of Hunan Provincial Education Department under Grants No. 22A0319.

\bibliographystyle{JHEP}
\bibliography{references}

@article{Wang:2022fih,
    author = "Wang, Jian-Peng and Qin, Qin and Yu, Fu-Sheng",
    title = "{Complementary CP violation induced by T-odd and T-even correlations}",
    eprint = "2211.07332",
    archivePrefix = "arXiv",
    primaryClass = "hep-ph",
    doi = "10.1103/cdgq-tbww",
    journal = "Phys. Rev. D",
    volume = "111",
    number = "11",
    pages = "L111301",
    year = "2025"
}

@article{Shi:2019vus,
    author = "Shi, Xiao-Dong and Kang, Xian-Wei and Bigi, Ikaros and Wang, Wei-Ping and Peng, Hai-Ping",
    title = "{Prospects for CP and P violation in $\Lambda_{c}^+$ decays at Super Tau Charm Facility}",
    eprint = "1904.12415",
    archivePrefix = "arXiv",
    primaryClass = "hep-ph",
    doi = "10.1103/PhysRevD.100.113002",
    journal = "Phys. Rev. D",
    volume = "100",
    number = "11",
    pages = "113002",
    year = "2019"
}

@article{Gronau:2015gha,
    author = "Gronau, Michael and Rosner, Jonathan L.",
    title = "{Triple product asymmmetries in $\Lambda_b$ and $\Xi_b$ decays}",
    eprint = "1506.01346",
    archivePrefix = "arXiv",
    primaryClass = "hep-ph",
    reportNumber = "EFI-15-18, TECHNION-PH-2015-07",
    doi = "10.1016/j.physletb.2015.07.060",
    journal = "Phys. Lett. B",
    volume = "749",
    pages = "104--107",
    year = "2015"
}

@article{Bensalem:2002ys,
    author = "Bensalem, Wafia and Datta, Alakabha and London, David",
    title = "{New physics effects on triple product correlations in Lambda(b) decays}",
    eprint = "hep-ph/0208054",
    archivePrefix = "arXiv",
    reportNumber = "UDEM-GPP-TH-02-102",
    doi = "10.1103/PhysRevD.66.094004",
    journal = "Phys. Rev. D",
    volume = "66",
    pages = "094004",
    year = "2002"
}

@article{Bensalem:2002pz,
    author = "Bensalem, Wafia and Datta, Alakabha and London, David",
    title = "{T violating triple product correlations in charmless Lambda(b) decays}",
    eprint = "hep-ph/0205009",
    archivePrefix = "arXiv",
    reportNumber = "UDEM-GPP-TH-02-96",
    doi = "10.1016/S0370-2693(02)02028-2",
    journal = "Phys. Lett. B",
    volume = "538",
    pages = "309--320",
    year = "2002"
}

@article{Bensalem:2000hq,
    author = "Bensalem, Wafia and London, David",
    title = "{$T$ odd triple product correlations in hadronic $b$ decays}",
    eprint = "hep-ph/0005018",
    archivePrefix = "arXiv",
    reportNumber = "UDEM-GPP-TH-00-70",
    doi = "10.1103/PhysRevD.64.116003",
    journal = "Phys. Rev. D",
    volume = "64",
    pages = "116003",
    year = "2001"
}

@article{Durieux:2015zwa,
    author = "Durieux, Gauthier and Grossman, Yuval",
    title = "{Probing CP violation systematically in differential distributions}",
    eprint = "1508.03054",
    archivePrefix = "arXiv",
    primaryClass = "hep-ph",
    reportNumber = "CP3-15-24",
    doi = "10.1103/PhysRevD.92.076013",
    journal = "Phys. Rev. D",
    volume = "92",
    number = "7",
    pages = "076013",
    year = "2015"
}

@article{Dunietz:1990cj,
    author = "Dunietz, Isard and Quinn, Helen R. and Snyder, A. and Toki, W. and Lipkin, Harry J.",
    title = "{How to extract CP violating asymmetries from angular correlations}",
    reportNumber = "SLAC-PUB-5270",
    doi = "10.1103/PhysRevD.43.2193",
    journal = "Phys. Rev. D",
    volume = "43",
    pages = "2193--2208",
    year = "1991"
}

@article{Kayser:1989vw,
    author = "Kayser, Boris",
    editor = "Singer, Paul and Eilam, Gad",
    title = "{Kinematically Nontrivial CP Violation in Beauty Decay}",
    reportNumber = "Print-89-0792 (NSF)",
    doi = "10.1016/0920-5632(90)90113-9",
    journal = "Nucl. Phys. B Proc. Suppl.",
    volume = "13",
    pages = "487--490",
    year = "1990"
}

@article{Valencia:1988it,
    author = "Valencia, German",
    title = "{Angular Correlations in the Decay $B \to V V$ and {CP} Violation}",
    reportNumber = "BNL-42229",
    doi = "10.1103/PhysRevD.39.3339",
    journal = "Phys. Rev. D",
    volume = "39",
    pages = "3339",
    year = "1989"
}

@article{Donoghue:1987wu,
    author = "Donoghue, John F. and Holstein, Barry R. and Valencia, German",
    title = "{Survey of Present and Future Tests of {CP} Violation}",
    reportNumber = "UMHEP-272",
    doi = "10.1142/S0217751X87000144",
    journal = "Int. J. Mod. Phys. A",
    volume = "2",
    pages = "319",
    year = "1987"
}

@article{Zhang:2022emj,
    author = "Zhang, Zhen-Hua and Qi, Jing-Juan",
    title = "{Analysis of angular distribution asymmetries and the associated $C\!P$ asymmetries in three-body decays of bottom baryons}",
    eprint = "2208.13411",
    archivePrefix = "arXiv",
    primaryClass = "hep-ph",
    doi = "10.1140/epjc/s10052-023-11267-7",
    journal = "Eur. Phys. J. C",
    volume = "83",
    number = "2",
    pages = "133",
    year = "2023"
}

@article{Zhang:2025mne,
    author = "Zhang, Zhen-Hua and Yang, Jian-Yu and Guo, Xin-Heng",
    title = "{Full analysis of CP violation induced by the decay angular correlations in four-body cascade decays of heavy hadrons}",
    eprint = "2504.19228",
    archivePrefix = "arXiv",
    primaryClass = "hep-ph",
    month = "4",
    year = "2025"
}

@article{BESIII:2018cnd,
    author = "Ablikim, M. and others",
    collaboration = "BESIII",
    title = "{Polarization and Entanglement in Baryon-Antibaryon Pair Production in Electron-Positron Annihilation}",
    eprint = "1808.08917",
    archivePrefix = "arXiv",
    primaryClass = "hep-ex",
    doi = "10.1038/s41567-019-0494-8",
    journal = "Nature Phys.",
    volume = "15",
    pages = "631--634",
    year = "2019"
}

@article{BESIII:2022qax,
    author = "Ablikim, M. and others",
    collaboration = "BESIII",
    title = "{Precise Measurements of Decay Parameters and $CP$ Asymmetry with Entangled $\Lambda-\bar{\Lambda}$ Pairs}",
    eprint = "2204.11058",
    archivePrefix = "arXiv",
    primaryClass = "hep-ex",
    doi = "10.1103/PhysRevLett.129.131801",
    journal = "Phys. Rev. Lett.",
    volume = "129",
    number = "13",
    pages = "131801",
    year = "2022"
}

@article{Cabibbo:1963yz,
    author = "Cabibbo, N.",
    title = "{Unitary Symmetry and Leptonic Decays}",
    doi = "10.1103/PhysRevLett.10.531",
    journal = "Phys. Rev. Lett.",
    volume = "10",
    pages = "531--533",
    year = "1963"
}

@article{Kobayashi:1973fv,
    author = "Kobayashi, M. and Maskawa, T.",
    title = "{CP Violation in the Renormalizable Theory of Weak Interaction}",
    reportNumber = "KUNS-242",
    doi = "10.1143/PTP.49.652",
    journal = "Prog. Theor. Phys.",
    volume = "49",
    pages = "652--657",
    year = "1973"
}

@article{Christenson:1964fg,
    author = "Christenson, J. H. and Cronin, J. W. and Fitch, V. L. and Turlay, R.",
    title = "{Evidence for the $2\pi$ Decay of the $K_2^0$ Meson}",
    doi = "10.1103/PhysRevLett.13.138",
    journal = "Phys. Rev. Lett.",
    volume = "13",
    pages = "138--140",
    year = "1964"
}

@article{KTeV:1999kad,
    author = "Alavi-Harati, A. and others",
    collaboration = "KTeV",
    title = "{Observation of Direct CP Violation in $K_{S,L} \to \pi \pi$ Decays}",
    eprint = "hep-ex/9905060",
    archivePrefix = "arXiv",
    reportNumber = "EFI-99-25, FERMILAB-PUB-99-150-E",
    doi = "10.1103/PhysRevLett.83.22",
    journal = "Phys. Rev. Lett.",
    volume = "83",
    pages = "22--27",
    year = "1999"
}

@article{BaBar:2001pki,
    author = "Aubert, B. and others",
    collaboration = "BaBar",
    title = "{Observation of CP violation in the $B^0$ meson system}",
    eprint = "hep-ex/0107013",
    archivePrefix = "arXiv",
    reportNumber = "SLAC-PUB-8904, BABAR-PUB-01-18",
    doi = "10.1103/PhysRevLett.87.091801",
    journal = "Phys. Rev. Lett.",
    volume = "87",
    pages = "091801",
    year = "2001"
}

@article{Belle:2001zzw,
    author = "Abe, K. and others",
    collaboration = "Belle",
    title = "{Observation of large CP violation in the neutral $B$ meson system}",
    eprint = "hep-ex/0107061",
    archivePrefix = "arXiv",
    reportNumber = "KEK-PREPRINT-2001-50, BELLE-PREPRINT-2001-10",
    doi = "10.1103/PhysRevLett.87.091802",
    journal = "Phys. Rev. Lett.",
    volume = "87",
    pages = "091802",
    year = "2001"
}

@article{Belle:2010xyn,
    author = "Poluektov, A. and others",
    collaboration = "Belle",
    title = "{Evidence for direct CP violation in the decay $B^\pm \rightarrow D^{(*)}K^\pm$, $D\rightarrow K^0_S\pi^+\pi^-$ and measurement of the CKM phase $\phi_3$}",
    eprint = "1003.3360",
    archivePrefix = "arXiv",
    primaryClass = "hep-ex",
    reportNumber = "BELLE-PREPRINT-2010-4",
    doi = "10.1103/PhysRevD.81.112002",
    journal = "Phys. Rev. D",
    volume = "81",
    pages = "112002",
    year = "2010"
}

@article{BaBar:2010hvw,
    author = "del Amo Sanchez, P. and others",
    collaboration = "BaBar",
    title = "{Measurement of CP observables in $B^\pm \rightarrow D_{CP}K^\pm$ decays and constraints on the CKM angle $\gamma$}",
    eprint = "1007.0504",
    archivePrefix = "arXiv",
    primaryClass = "hep-ex",
    reportNumber = "SLAC-PUB-14187, BABAR-PUB-10-008",
    doi = "10.1103/PhysRevD.82.072004",
    journal = "Phys. Rev. D",
    volume = "82",
    pages = "072004",
    year = "2010"
}

@article{LHCb:2013syl,
    author = "Aaij, R. and others",
    collaboration = "LHCb",
    title = "{First observation of $CP$ violation in the decays of $B^0_s$ mesons}",
    eprint = "1304.6173",
    archivePrefix = "arXiv",
    primaryClass = "hep-ex",
    reportNumber = "CERN-PH-EP-2013-068, LHCB-PAPER-2013-018",
    doi = "10.1103/PhysRevLett.110.221601",
    journal = "Phys. Rev. Lett.",
    volume = "110",
    number = "22",
    pages = "221601",
    year = "2013"
}

@article{LHCb:2019hro,
    author = "Aaij, R. and others",
    collaboration = "LHCb",
    title = "{Observation of CP Violation in Charm Decays}",
    eprint = "1903.08726",
    archivePrefix = "arXiv",
    primaryClass = "hep-ex",
    reportNumber = "LHCb-PAPER-2019-006, CERN-EP-2019-042",
    doi = "10.1103/PhysRevLett.122.211803",
    journal = "Phys. Rev. Lett.",
    volume = "122",
    number = "21",
    pages = "211803",
    year = "2019"
}

@article{LHCb:2025ray,
    author = "Aaij, R. and others",
    collaboration = "LHCb",
    title = "{Observation of charge{\textendash}parity symmetry breaking in baryon decays}",
    eprint = "2503.16954",
    archivePrefix = "arXiv",
    primaryClass = "hep-ex",
    reportNumber = "LHCb-PAPER-2024-054, CERN-EP-2025-031",
    doi = "10.1038/s41586-025-09119-3",
    journal = "Nature",
    volume = "643",
    number = "8074",
    pages = "1223--1228",
    year = "2025"
}

@article{Lee:1957qs,
    author = "Lee, T. D. and Yang, C. N.",
    title = "{General Partial Wave Analysis of the Decay of a Hyperon of Spin 1/2}",
    doi = "10.1103/PhysRev.108.1645",
    journal = "Phys. Rev.",
    volume = "108",
    pages = "1645--1647",
    year = "1957"
}

@article{Donoghue:1986hh,
    author = "Donoghue, J. F. and He, X. G. and Pakvasa, S.",
    title = "{Hyperon Decays and CP Nonconservation}",
    reportNumber = "UH-511-579-86, UMHEP-245",
    doi = "10.1103/PhysRevD.34.833",
    journal = "Phys. Rev. D",
    volume = "34",
    pages = "833",
    year = "1986"
}

@article{Zhang:2021fdd,
    author = "Zhang, Z. H. and Guo, X. H.",
    title = "{A novel strategy for searching for CP violations in the baryon sector}",
    eprint = "2103.11335",
    archivePrefix = "arXiv",
    primaryClass = "hep-ph",
    doi = "10.1007/JHEP07(2021)177",
    journal = "JHEP",
    volume = "07",
    pages = "177",
    year = "2021"
}

@article{LHCb:2019sus,
    author = "Aaij, R. and others",
    collaboration = "LHCb",
    title = "{Amplitude analysis of the $B^+ \rightarrow \pi^+\pi^+\pi^-$ decay}",
    eprint = "1909.05212",
    archivePrefix = "arXiv",
    primaryClass = "hep-ex",
    reportNumber = "LHCb-PAPER-2019-017, CERN-EP-2019-157",
    doi = "10.1103/PhysRevD.101.012006",
    journal = "Phys. Rev. D",
    volume = "101",
    number = "1",
    pages = "012006",
    year = "2020"
}

\end{document}